\documentclass[aps,prl,twocolumn,showpacs,superscriptaddress,10pt]{revtex4-1}

\usepackage{amsmath,amsfonts,amssymb}
\usepackage{graphicx}
\usepackage{hyperref}
\usepackage{tensor}
\usepackage{multirow}
\usepackage{nicefrac}
\usepackage{cases}
\usepackage[vcentermath]{youngtab}
\usepackage{xspace}

\usepackage{slashed}
\usepackage{subfig}
 
\usepackage{latexsym,epsfig}

\usepackage[final]{showkeys}
\usepackage{amsfonts}
\usepackage{mathrsfs}
\usepackage{dsfont}
\usepackage[Symbol]{upgreek}

\usepackage{tcolorbox}
\usepackage{empheq}
\usepackage{cancel}

\newcommand{\nn}{\nonumber}

\newcommand{\be}{\begin{align}}
\newcommand{\ee}{\end{align}}

\newcommand{\bea}{\begin{eqnarray}}
\newcommand{\eea}{\end{eqnarray}}

\def\*{\partial}

\def\={\!=\!}

\usepackage{color}
\definecolor{red}{rgb}{1,0,0}
\definecolor{lred}{rgb}{0.3,0,0}
\definecolor{green}{rgb}{0,0.6,0}
\definecolor{blue}{rgb}{0,0,1}
\definecolor{violet}{rgb}{0.8,0,0.8}

\definecolor{darkred}{rgb}{0.65,0.15,0}
\definecolor{darkgreen}{rgb}{.05,.5,.05}
\hypersetup{pdfborder={0 0 0},colorlinks=true,urlcolor=darkred,citecolor=darkred,linkcolor=darkred,linktocpage=true}

\newcommand{\SO}{\ensuremath{\mathrm{SO}}\xspace}
\newcommand{\Odd}{\ensuremath{\mathrm{O}(d,d,\mathbb{R})}\xspace}
\newcommand{\odd}{\ensuremath{\mathfrak{o}(d,d,\mathbb{R})}\xspace}

\newcommand{\GL}{\ensuremath{\mathrm{GL}}\xspace}

\newcommand{\F}{\ensuremath{\mathcal{F}}\xspace}
\renewcommand{\S}{\ensuremath{\mathcal{S}}\xspace}
\renewcommand{\H}{\ensuremath{\mathcal{H}}\xspace}

\renewcommand{\d}{\ensuremath{\mathrm{d}}\xspace}
\newcommand{\Tr}[1]{\ensuremath{\mathrm{Tr}\left(#1\right)}\xspace}

\begin{document}

\title{Green-Schwarz Mechanism for String Dualities}
\author{Camille Eloy}
\affiliation{Univ Lyon, Ens de Lyon, Univ Claude Bernard, CNRS,
Laboratoire de Physique, F-69342 Lyon, France}
\author{Olaf Hohm}
\affiliation{Institute for Physics, Humboldt University Berlin, Zum Grossen Windkanal 6, D-12489 Berlin, Germany}
\author{Henning Samtleben}
\affiliation{Univ Lyon, Ens de Lyon, Univ Claude Bernard, CNRS,
Laboratoire de Physique, F-69342 Lyon, France}

\date{\today}

\begin{abstract}

We determine the complete spacetime action to first order in $\alpha'$ for the massless fields of bosonic string theory compactified on a $d$-dimensional torus. 
A fully systematic procedure is developed that brings the action into a minimal form in which all fields apart from the metric enter only with first-order derivatives. 
T-duality implies that this action must have a global \Odd symmetry, and we show 
that this  requires a  Green-Schwarz type mechanism  for $\alpha'$-deformed \Odd transformations. 
In terms of a frame formalism with $\GL(d)\times \GL(d)$ gauge symmetry this amounts to a modification of the three-form curvature by 
a Chern-Simons term for composite gauge fields.

\end{abstract}

\maketitle


String theory is a  particularly promising candidate for a consistent theory  of quantum gravity, 
but it is fair to say that its underlying principles remain elusive. 
An important clue is that, even classically, string theory 
modifies general relativity in two significant ways. 
First, string theory features novel \textit{dualities} which imply that theories 
defined on seemingly different backgrounds are actually equivalent. 
Second, the classical field equations receive an infinite number of higher-derivative 
corrections governed by the dimensionful (inverse) string tension $\alpha'$. 
Whatever the fundamental formulation of string or M-theory may be, 
it seems clear that it would have to accommodate these two features as core principles. 
In this Letter 
we report on results for a fully systematic procedure to determine the duality invariant spacetime action for massless string fields 
to higher order in $\alpha'$ and  point out a curious interplay between string dualities and $\alpha'$ corrections. 
We discuss only the main results; the technical details will be presented in Ref.~\cite{EloyHohmSamtleben}.

The simplest  duality shared by all closed string theories is T-duality. It 
states that string backgrounds containing a $d$-dimensional torus
are mapped under the group ${\rm O}(d,d,\mathbb{Z})$  to physically equivalent backgrounds. 
This duality includes, for a single circle, the inversion of the radius $R\rightarrow \alpha'/R$. 
The T-duality property of closed string theory implies that the spacetime action for the massless 
string fields on such backgrounds features a global \Odd symmetry \cite{Sen:1991zi}. 
To lowest order in $\alpha'$ this symmetry was first displayed in the cosmological setting (reduction to one dimension) in Ref.~\cite{Veneziano:1991ek} 
and later generalized to arbitrary $d$ in Ref.~\cite{Maharana:1992my}. 
Subsequent seminal work by Meissner revealed an \Odd invariance to first order in $\alpha'$ in the cosmological setting, 
with the group action being undeformed thanks to a judicious choice of field variables \cite{Meissner:1996sa}. In Ref.~\cite{Kaloper:1997ux} 
the case of a single circle was investigated, while Ref.~\cite{Godazgar:2013bja} includes a general torus but truncates to `internal' field degrees of freedom.

More recently, $\alpha'$ corrections have been investigated in the extended framework of double field theory \cite{Siegel:1993th,Hull:2009mi,Hohm:2010pp},
which is the duality-covariant formulation of spacetime actions before compactification. 
Notably,  two unexpected new features appear: 1) the gauge transformations need to be $\alpha'$-corrected, and 2) 
there is no background independent formulation in terms of the familiar \Odd matrix encoding metric and $B$-field \cite{Hohm:2013jaa,Hohm:2014eba,Hohm:2014xsa,Marques:2015vua,Hohm:2016lge,Hohm:2016yvc}. 
Rather, the general formulation of double field theory to first order in $\alpha'$ 
employs a frame formalism 
with $\alpha'$ corrected tangent space transformations. 
While in the dimensionally reduced theories determined so far 
there is a choice of field basis for which the \Odd action is undeformed at order $\alpha'$, it has remained an open 
question whether $\alpha'$-deformations of \Odd may be required in general dimensionally reduced theories. 
The reduction of $\alpha'$-deformed double field theory has been investigated in Ref.~\cite{Baron:2017dvb}, 
however, without extracting the consequences for the realization of \Odd in the 
dimensional reduction of conventional (non-extended) theories.

In this Letter we give the complete action to first order in $\alpha'$ for arbitrary  $d$ and 
show that in general the \Odd transformations are $\alpha'$-deformed according to a novel Green-Schwarz type mechanism.  
The original Green-Schwarz mechanism, which triggered the first superstring revolution, is needed to show that gravitational and gauge  
anomalies can be cancelled, for gauge groups $\SO(32)$ or ${\rm E}_8\times {\rm E}_8$, by an $\alpha'$-deformation of the gauge transformations 
of the (singlet) $B$-field, thereby modifying the classical (tree-level) theory~\cite{Green:1984sg}. 
Similarly, we show here that \Odd invariance of the spacetime action reduced on a $d$-dimensional torus requires 
a non-trivial transformation for the singlet $B$-field already classically, with a Chern-Simons type modification of the  3-form curvature. 
This suggests that Green-Schwarz type mechanisms may be much more ubiquitous than expected.

We consider the two-derivative effective action for the bosonic string 
in $D+d$ dimensions, with metric $\hat{g}_{\hat\mu\hat\nu}$, 
antisymmetric Kalb-Ramond field $\hat B_{\hat\mu\hat\nu}$ and the dilaton~$\hat\phi$
\begin{equation}
\label{eq:I0D+d}
\hat{I}_{0} =\!\! \int\! \d^{D+d}X\,\sqrt{-\hat{g}}\,e^{-\hat{\phi}}\,\left(\hat{R} + \partial_{\hat\mu}\hat\phi\,\partial^{\hat\mu}\hat\phi-\frac{1}{12} \hat H^{2}\right), 
\end{equation}
where indices $\hat\mu$ run over the $(D+d)$ dimensional space, and 
$\hat{H}^{2}= \hat H^{\hat\mu\hat\nu\hat\rho}\hat H_{\hat\mu\hat\nu\hat\rho}$ with the field strength
$\hat H_{\hat\mu\hat\nu\hat\rho}=3\,\partial_{[\hat\mu}\hat B_{\hat\nu\hat\rho]}$. For compactification on the $d$-dimensional torus, we use the index split 
\bea
X^{\hat\mu}=(x^{\mu},y^{m})\;,\;\;\mu\in[\![1,D]\!]\,,\;\;
m\in[\![1,d]\!]\;,
\eea 
and drop the field dependance on the internal coordinates~$y^{m}$. For the metric $\hat{g}_{\hat\mu\hat\nu}$, we consider the standard Kaluza-Klein ansatz
 \begin{equation}
 \label{eq:ansatzg}
 \hat{g}_{\hat{\mu}\hat{\nu}}= \begin{pmatrix}
                                      g_{\mu\nu}+A_{\mu}^{(1)p}G_{pq}A_{\nu}^{(1)q} & A_{\mu}^{(1)p}G_{pn} \\[.5ex]
                                      G_{mp}A_{\nu}^{(1)p} & G_{mn}
                              \end{pmatrix}
                              \;,
                              \medskip\\
 \end{equation}
in terms of the $D$-dimensional metric $g_{\mu\nu}$, Kaluza-Klein vector fields $A_{\mu}^{(1)m}$ and the internal metric $G_{mn}$. 
Similarly, for the 2-form $\hat{B}_{\hat\mu\hat\nu}$, we use the ansatz
\begin{equation}
\label{eq:ansatzB}
\begin{cases}
\hat{B}_{mn} =B_{mn}\;,\\
\hat{B}_{\mu m} = A^{(2)}_{\mu\,m}-A_{\mu}^{(1)n}B_{mn}\;, \\
\hat{B}_{\mu\nu} = B_{\mu\nu}-A_{[\mu}^{(1)m}A_{\nu]m}^{(2)}+A_{\mu}^{(1)m} \,B_{mn}\,A_{\nu}^{(1)n}\;, 
\end{cases}
\end{equation}
in terms of $D$-dimensional scalars $B_{mn}$, vector fields $A_{\mu\,m}^{(2)}$, and a 2-form $B_{\mu\nu}$. 
After dimensional reduction, the action (\ref{eq:I0D+d}) may be cast into the manifestly 
\Odd invariant form~\cite{Maharana:1992my}
\begin{align}
I_{0}=&\int\d^{D}x\,\sqrt{-g}\,e^{-\Phi}\,\Big(R+\partial_{\mu}\Phi\,\partial^{\mu}\Phi-\frac{1}{12} H^{2} 
\nonumber\\
&
+\frac{1}{8}\Tr{\nabla_{\mu}\H \nabla^{\mu}\H^{-1}}-\frac{1}{4}\F_{\mu\nu}{}^{M} \H_{M N} \F^{\mu\nu\,N}\Big)\,. 
\label{eq:MaharanaSchwarz}
\end{align} 
Here, $\Phi = \hat{\phi}-\dfrac{1}{2}\,\log(\det(G_{mn}))$ is the shifted dilaton and the \Odd matrix ${\cal H}$ is given by
\begin{equation}
  {\cal H}_{MN} = \begin{pmatrix}
  G_{mn}-B_{mp}G^{pq}B_{qn} & B_{mp}G^{pn}\\[.5ex]
                    -G^{mp}B_{pn} & G^{mn}                 
                    \end{pmatrix}\;.
                    \label{HOdd}
\end{equation}
 The vectors $A^{(1)m}_{\mu}$ and $A^{(2)}_{\mu\,m}$ are combined into an \Odd vector
\begin{equation}
 {\cal A}_{\mu}{}^{M} = \begin{pmatrix}
                        A_{\mu}^{(1)\,m} \\
                        A_{\mu\,m}^{(2)}
                        \end{pmatrix}\;
                        \label{AOdd},
\end{equation}
and $\F_{\mu\nu}{}^{M}=2\partial_{[\mu}\cal A_{\nu]}{}^{M}$ is the associated abelian field-strength. 
Finally, the $D$-dimensional 3-form $H_{\mu\nu\rho}$ in Eq.~\eqref{eq:MaharanaSchwarz} is given by
\begin{equation}
H_{\mu\nu\rho} = 3\left(\partial_{[\mu}B_{\nu\rho]}-\frac{1}{2}{\cal A}_{[\mu}{}^{M}\F_{\nu\rho]\,M}\right)\,.
\label{HAA}
\end{equation}
Throughout, the \Odd indices are raised and lowered using the constant \Odd-invariant matrix
\bea\label{etamatrix}
  \eta^{MN} = \begin{pmatrix}
                    0 & {\delta^{m}}_{n} \\
                    {\delta_{m}}^{n} & 0
                  \end{pmatrix}\;.
\eea
In terms of the covariant objects (\ref{HOdd}), (\ref{AOdd}),
the action~\eqref{eq:MaharanaSchwarz} makes the invariance under \Odd transformations manifest. 
The goal of this Letter is the extension of this construction in presence of
higher order corrections at first order in $\alpha'$.

At first order in $\alpha'$, the correction to the action~\eqref{eq:I0D+d} is given by~\cite{Metsaev:1987zx}
\begin{align}
\label{eq:alpha'D+d}
&\hat{I}_{1}=
\frac{\alpha'}{4} \!\!\int\!\d^{D+d}X\sqrt{-\hat{g}}\,e^{-\hat{\phi}}\Big(\hat{R}_{\hat\mu\hat\nu\hat\rho\hat\sigma}\hat{R}^{\hat\mu\hat\nu\hat\rho\hat\sigma}-\frac{1}{8}\,\hat{H}^{2}_{\hat\mu \hat\nu}\hat{H}^{2\,\hat\mu \hat\nu}\nonumber \\
&
-\frac{1}{2} \hat{H}^{\hat\mu\hat\nu\hat\lambda}{\hat{H}^{\hat\rho\hat\sigma}}_{\ \ \,\hat\lambda}\, \hat{R}_{\hat\mu\hat\nu\hat\rho\hat\sigma} 
+\frac{1}{24}\hat{H}_{\hat\mu \hat\nu \hat\rho}\hat{H}^{\hat\mu\ \hat\lambda}_{\ \,\hat\sigma}\hat{H}^{\hat\nu\ \hat\tau}_{\ \,\hat\lambda}\hat{H}^{\hat\rho\ \hat\sigma}_{\ \,\hat\tau}\Big)\;,
\end{align}
up to field redefinitions. Here, $\hat{R}_{\hat\mu\hat\nu\hat\rho\hat\sigma}$ is the Riemann tensor, and 
we defined  $\hat{H}^{2}_{\hat\mu \hat\nu} \equiv \hat{H}^{\hat\mu \hat\rho\hat\sigma}\hat{H}^{\hat\nu}_{\ \,\hat\rho\hat\sigma}$. 
In the following, we compactify this action on the $d$-dimensional torus and seek to bring it into a manifestly \Odd invariant form.
In the presence of higher order corrections this construction appears far less straightforward than in the two-derivative case due to the many potential
ambiguities from field redefinitions and partial integrations.

Let us outline our systematics, which generalize those in Refs.~\cite{Hohm:2015doa,Hohm:2019jgu} to arbitrary dimensions. 
We first determine a basis of independent \Odd invariant four-derivative terms.
In $D$ dimensions, every four-derivative term carrying the leading two-derivative contribution from the field equations
descending from the two-derivative action (\ref{eq:MaharanaSchwarz}) can be eliminated by a field redefinition in order $\alpha'$.
Moreover, different four-derivative terms in the Lagrangian can be related by integration by parts.
Upon a systematic count, dividing out this freedom,
we find that at order $\alpha'$, there are 61 independent four-derivative
terms that can be built from the \Odd covariant objects 
${\cal H}_{MN}$, ${\cal F}_{\mu\nu}{}^M$, $H_{\mu\nu\rho}$, $R_{\mu\nu\rho\sigma}$ and their derivatives. 
By explicit construction, we then show the existence of a distinguished basis, which, except for the $D$-dimensional Riemann tensor, only
carries first order derivative terms, i.e. which is polynomial in $\nabla_{\mu}\Phi$, $\nabla_\mu {\cal H}_{MN}$, ${\cal F}_{\mu\nu}{}^M$, $H_{\mu\nu\rho}$, and $R_{\mu\nu\rho\sigma}$.

Next, we compactify separately every term from Eq.~\eqref{eq:alpha'D+d} with the ansatz (\ref{eq:ansatzg}), (\ref{eq:ansatzB}) and find that all terms
carrying second order derivatives can indeed be eliminated by field redefinitions and  integrations by part. Generically, these field redefinitions will
not be \Odd covariant. The remaining terms then can be matched
against the previously determined 61-dimensional  
\Odd invariant basis after breaking the terms of the latter under ${\rm GL}(d)$ according to Eqs.~\eqref{HOdd} and \eqref{AOdd}.
Details of this rather lengthy computation will appear in Ref.~\cite{EloyHohmSamtleben}.
As a result, we find that after compactification the action (\ref{eq:alpha'D+d}) can be brought into the form
\bea
\hat{I}_{1} &\longrightarrow& 
I_{1} + O_1
\;,
\eea
where $I_{1}$ takes the manifestly \Odd invariant form
\begin{widetext}
  \begin{align}
  \label{eq:alpha'D}
  I_{1} &= \frac{1}{4}\,\alpha' \int\d^{D}x\,\sqrt{-g}\,e^{-\Phi}\,\Big[R_{\mu\nu\rho\sigma}R^{\mu\nu\rho\sigma}-\frac{1}{2}\, R_{\mu\nu\rho\sigma}{H}^{\mu\nu\lambda}{H}^{\rho\sigma}_{\ \ \,\lambda}-\frac{1}{8}\,{H}^{2}_{\mu\nu}{H}^{2\,\mu\nu} + \frac{1}{24}\, {H}_{\mu\nu\rho}{H}^{\mu\ \, \lambda}_{\ \,\sigma}{H}^{\nu\ \,\tau}_{\ \,\lambda}{H}^{\rho\ \,\sigma}_{\ \,\tau} \nonumber \\
  &+\frac{1}{16}\,\Tr{\nabla_{\mu}\S\nabla_{\nu}\S\nabla^{\mu}\S\nabla^{\nu}\S} - \frac{1}{32}\, \Tr{\nabla_{\mu}\S\nabla_{\nu}\S} \Tr{\nabla^{\mu}\S\nabla^{\nu}\S}+\frac{1}{8}\, {\F_{\mu\nu}}^{M}{\S_{M}}^{N}\F_{\rho\sigma\,N}\F^{\mu\rho\,P}{\S_{P}}^{Q}{\F^{\nu\sigma}}_{Q} \nonumber \\
  &-\frac{1}{2}\,{\F_{\mu\nu}}^{M}{\S_{M}}^{N}{\F^{\mu\rho}}_{N}\F^{\nu\sigma\,P}{\S_{P}}^{Q}\F_{\rho\sigma\,Q}+\frac{1}{8}\,{\F_{\mu\nu}}^{M}{\F_{\rho\sigma}}_{M}\F^{\mu\rho\,N}{\F^{\nu\sigma}}_{N}-\frac{1}{2}\, R_{\mu\nu\rho\sigma} \F^{\mu\nu\,M}{\S_{M}}^{N}{\F^{\rho\sigma}}_{N} \\
  &+\frac{1}{8}\,{H}^{2}_{\mu\nu}\Tr{\nabla^{\mu}\S\nabla^{\nu}\S}-\frac{1}{2}\,{H}^{2}_{\mu\nu}{{\F^{\mu}}_{\rho}}^{M}{\S_{M}}^{N}{\F^{\nu\rho}}_{N} +\frac{1}{4}\, {H}^{\mu\nu\lambda}{H}^{\rho\sigma}_{\ \ \,\lambda}{\F_{\mu\rho}}^{M}{\S_{M}}^{N}\F_{\nu\sigma\,N} \nn\\
  &-\frac{1}{2}\,{\F_{\mu\nu}}^{M}{\left(\S\nabla_{\rho}\S\nabla^{\nu}\S\right)_{M}}^{N} {\F^{\mu\rho}}_{N} + \frac{1}{4}\,\F^{\mu\rho\,M}{\S_{M}}^{N}{\F^{\nu}}_{\rho\,N}\Tr{\nabla_{\mu}\S\nabla_{\nu}\S}-\frac{1}{2}\,{H}^{\mu\nu\rho}{\F_{\mu\sigma}}^{M}{\left(\S\nabla_{\nu}\S\right)_{M}}^{N}{{\F_{\rho}}^{\sigma}}_{N} \Big]\;, \nn
  \end{align}
\end{widetext}
with ${H}^{2}_{\mu \nu} = {H}^{\mu \rho\sigma}{H}^{\nu}_{\ \,\rho\sigma}$, and
the matrix ${\cal S}$ defined as ${\S_{M}}^{N} = {\cal H}_{MP}\,\eta^{PN}$. 
In contrast, the term $O_1$ is not \Odd invariant, but
takes the particular form
\bea
O_1 &=& \frac16\,\alpha' \int\d^{D}x\,\sqrt{-g}\,e^{-\Phi}\,H_{\mu\nu\rho}\,\Omega^{\mu\nu\rho}
\;,
\label{eq:Itilde}
\eea
with $\Omega_{\mu\nu\rho}$ given  by 
\begin{align}\label{Omegafirst}
\Omega_{\mu\nu\rho} = &-\frac{3}{4}\,{\rm Tr}\big(\partial_{[\mu}G^{-1} G \partial_{\nu}G^{-1} \partial_{\rho]} B\big)\nn\\
&+\frac{1}{4}\,{\rm Tr}\big(\partial_{[\mu}B G^{-1} \partial_{\nu}B G^{-1}\partial_{\rho]}B G^{-1}\big)\;.
\end{align}
This 3-form has a remarkable structure owing its existence to the non-vanishing cohomology $H^4(M)$ of
$M=\Odd/({\rm O}(d)\times {\rm O}(d))$\,. 
Although $\Omega_{\mu\nu\rho}$ it is not \Odd invariant, its exterior derivative is,
as it takes the manifestly invariant form
\begin{equation}\label{Sinvariant}
4\,\partial_{[\mu}\Omega_{\nu\rho\sigma]}=\frac{3}{8}\,\Tr{\S\,\partial_{[\mu}\S\,\partial_{\nu}\S\,\partial_{\rho}\S\,\partial_{\sigma]}\S}
\;.
\end{equation}
We may thus conclude, that the \Odd variation of $\Omega_{\mu\nu\rho}$ is a closed 3-form,
which locally may be expressed as
\begin{equation}
\delta \Omega_{\mu\nu\rho} = 3\, \partial_{[\mu}X_{\nu\rho]}\;,
\label{eq:CStransform}
\end{equation}
in terms of a 2-form $X_{\mu\nu}$.
This observation together with the particular form of Eq.~\eqref{eq:Itilde}
suggests a Green-Schwarz type mechanism in oder to restore \Odd invariance of the $D$-dimensional action.

Namely, we may rewrite the result of the compactification of Eqs.~\eqref{eq:I0D+d} and \eqref{eq:alpha'D+d} as
\bea
\hat{I}_0 + \hat{I}_1 &\longrightarrow& \tilde{I}_0 + \tilde{I}_1 + {\cal O}(\alpha'{}^2)
\;,
\eea
where $\tilde{I}_0$, $\tilde{I}_1$ denote the 
actions (\ref{eq:MaharanaSchwarz}) and (\ref{eq:alpha'D}), respectively, in which 
we have replaced all 3-form curvatures $H_{\mu\nu\rho}$ by 
the deformation
  \begin{equation}
  \label{eq:newHH}
  \widetilde{H}_{\mu\nu\rho}\equiv H_{\mu\nu\rho} -\,\alpha'\,\Omega_{\mu\nu\rho}\;.
  \end{equation}
Indeed, this deformation of the lowest order action (\ref{eq:MaharanaSchwarz}) precisely produces  the non-invariant term 
(\ref{eq:Itilde}) at order $\alpha'$ while the deformation of Eq.~\eqref{eq:alpha'D} only contributes to higher orders in $\alpha'$\,.
As a last step, we observe that 
the deformed field strength~\eqref{eq:newHH} is O$(d,d,\mathbb{R})$ invariant if the 2-form $B_{\mu\nu}$ acquires a non-trivial variation
\begin{equation}
\delta B_{\mu\nu} = 
\alpha' X_{\mu\nu}
\;,
\label{eq:deltaB}
\end{equation}
under \Odd transformations, with the r.h.s.\ defined by integrating Eq.~\eqref{eq:CStransform}.
With this new anomalous transformation, the resulting theory is then fully \Odd invariant
to first order in $\alpha'$.
As expected, this is only true for the precise choice of relative coefficients
in the $(D+d)$-dimensional action (\ref{eq:alpha'D+d}).
In fact, the same method allows to prove a stronger result: 
starting from the most general action of eight independent
four-derivative terms in $(D+d)$ dimensions \cite{Metsaev:1987zx}, O$(d,d,\mathbb{R})$ invariance of the reduced action uniquely 
fixes all the coefficients up to an overall factor.

An explicit calculation of the variation~\eqref{eq:CStransform} for a general $ \odd$ generator
\begin{equation}
{\alpha_{M}}^{N} = \begin{pmatrix}
                      {\mathfrak{a}_{m}}^{n} & \mathfrak{b}_{m n} \\
                      \mathfrak{c}^{m n} & -{\mathfrak{a}_{n}}^{m}
                    \end{pmatrix} ,
\end{equation}
with $\mathfrak{c}^{mn}$ and $\mathfrak{b}_{mn}$ antisymmetric, yields the explicit form of the new \odd transformation of $B_{\mu\nu}$ as
\begin{equation}
X_{\mu\nu}= \frac{1}{2}\Tr{\mathfrak{c}\,\partial_{[\mu}(G+B)G^{-1}\partial_{\nu]}(G+B)}\;.
\label{eq:trafX}
\end{equation}
The 2-form thus acquires new transformations only along the nilpotent $\odd$ generators $\mathfrak{c}^{mn}$.
This is consistent with the fact that all the other $\odd$ generators have a geometric origin and
hence should be a manifest invariance of 
the dimensionally reduced action.
Moreover, with the expression (\ref{eq:trafX}), the algebra of \odd transformations closes on $B_{\mu\nu}$.
Crucially, the deformed \odd action~(\ref{eq:deltaB}) cannot be absorbed into a redefinition of the fields.

The action~\eqref{eq:alpha'D} is in agreement with previous results in the literature, up to field redefinitions. 
For $D=1$, it reproduces the result of Refs.~\cite{Meissner:1996sa,Hohm:2015doa}.
Once truncated to scalars only (i.e.\ setting ${\cal A}_\mu{}^M=0=B_{\mu\nu}$, $g_{\mu\nu}=\eta_{\mu\nu}$), 
the action (\ref{eq:alpha'D}) reduces to two terms which provide an equivalent compact rewriting of 
the action given in Ref.~\cite{Godazgar:2013bja}. 
However, the need for the Green-Schwarz type mechanism (\ref{eq:deltaB}) is not visible in any of these limits, 
as the mechanism mixes the Kalb-Ramond field $B_{\mu\nu}$ and the scalars $B_{mn}$.

It is instructive to inspect the above Green-Schwarz deformation in view of the $\mathbb{Z}_2$ invariance 
of bosonic string theory that sends $\hat{B}\rightarrow -\hat{B}$. On the \Odd matrix (\ref{HOdd}) this symmetry acts via the matrix $Z$ \cite{Hohm:2010pp}: 
 \begin{equation}\label{Z2metric}
  {\cal H} \rightarrow Z^T {\cal H} Z\,,\quad Z\equiv \begin{pmatrix}
                    1 & 0 \\
                    0 & -1
                  \end{pmatrix}. 
 \end{equation}
This matrix satisfies $Z=Z^{-1}=Z^T$ but is not  \Odd valued since the metric (\ref{etamatrix}) transforms as 
 \begin{equation}
  \eta\rightarrow Z\eta Z^T = -\eta\,.
 \end{equation}
Consequently, the matrix ${\cal S}$ obtained from ${\cal H}$ by raising of one index transforms as 
 \begin{equation}
  {\cal S}\rightarrow -Z{\cal S}Z\;. 
 \end{equation}
Thus, the \Odd invariant defined by the right-hand side of Eq.~(\ref{Sinvariant}) is $\mathbb{Z}_2$ odd. 
This is as it should be for $\mathbb{Z}_2$ invariance of the action since $B_{\mu\nu}$ and its field strength 
are also $\mathbb{Z}_2$ odd.

We have seen that the \textit{rigid} $O(d,d,\mathbb{R})$ transformations need to be $\alpha'$-deformed via a Green-Schwarz type mechanism. 
We now show that the theory can be reformulated, by means of a frame formalism, so that instead the gauge group of \textit{local} frame transformations gets 
$\alpha'$-deformed. This formulation then follows the standard Green-Schwarz mechanism more closely, albeit with composite gauge fields.

We introduce a frame field $E\equiv (E_{M}{}^{A})$ with inverse $E^{-1}\equiv (E_{A}{}^{M})$ in terms of which the 
matrix (\ref{HOdd})
is given by 
 \bea
  {\cal H}_{MN} = E_{M}{}^{A} E_{N}{}^{B}\kappa_{AB}\;, 
 \eea
where flat indices are split as $A=(a,\bar{a})$, and 
we assume $\kappa_{AB}$ to be block-diagonal with components $\kappa_{ab}$ and $\kappa_{\bar{a}\bar{b}}$. 
Since we make no further a priori assumption on $\kappa$ there is a local $\GL(d)\times \GL(d)$ frame invariance, 
acting as 
 \bea
  \delta_{\Lambda}E_{A}{}^{M} =\Lambda_{A}{}^{B} E_{B}{}^{M}\,, \quad \Lambda_{A}{}^{B}= \begin{pmatrix}  \Lambda_{a}{}^{b} & 0 \\[0.7ex]
 0 & \bar\Lambda_{\bar{a}}{}^{\bar{b}}
\end{pmatrix}. 
 \eea
Gauge fixing $\kappa_{AB}=\delta_{AB}$ the above reduces to the familiar frame formalism with local $\SO(d)\times \SO(d)$ 
invariance but for our present application an alternative gauge fixing is more convenient: using matrix notation, we set 
 \bea
 \kappa = \begin{pmatrix}  2G & 0 \\[0.7ex]
 0 & 2G
\end{pmatrix}\;, 
 \eea
and 
 \begin{equation}
 \label{sspecialgauge}
  E = \frac{1}{2} \begin{pmatrix}
                  1+BG^{-1} & 1-BG^{-1} \\[0.7ex]
                  G^{-1} & -G^{-1}
                  \end{pmatrix}\, .
 \end{equation}
Defining the Maurer-Cartan forms 
 \bea\label{MC}
  (E^{-1}\partial_{\mu}E)_A{}^{B} \equiv \begin{pmatrix}  Q_{\mu a}{}^{b} & P_{\mu a}{}^{\bar{b}} \\[0.7ex]
 \bar{P}_{\mu \bar{a}}{}^{b} & \bar{Q}_{\mu \bar{a}}{}^{\bar{b}}
\end{pmatrix}
 \eea
one finds that the $P_{\mu}$ transform as $\GL(d)\times \GL(d)$ tensors  while the $Q_{\mu}$ transform as $\GL(d)\times \GL(d)$ connections: 
 \bea\label{Qgauge}
  \delta_{\Lambda}Q_{\mu a}{}^{b} = -D_{\mu}\Lambda_{a}{}^{b}\;, \qquad  \delta_{\Lambda}\bar{Q}_{\mu \bar{a}}{}^{\bar{b}} = -D_{\mu}\bar{\Lambda}_{\bar{a}}{}^{\bar{b}}\;, 
 \eea
where $D_{\mu}\Lambda_a{}^{b}=\partial_{\mu}\Lambda_a{}^{b}+[Q_{\mu},\Lambda]_a{}^{b}$, and analogously  for the barred version. For the gauge choice (\ref{sspecialgauge})  the explicit form of these connections reads,
in matrix notation, 
  \begin{equation}
   \label{explicitQ}
  \begin{cases}
  Q_{\mu} = -\dfrac{1}{2}\partial_{\mu}(G-B)G^{-1}, \smallskip\\  
  \bar{Q}_{\mu} = -\dfrac{1}{2}\partial_{\mu}(G+B)G^{-1}\;. 
 \end{cases} 
  \end{equation}
We can now consider the Chern-Simons 3-form built from these connections: 
 \bea
  {\rm CS}_{\mu\nu\rho}(Q) \equiv {\rm Tr}\Big(Q_{[\mu}\partial_{\nu}Q_{\rho]}+\frac{2}{3}Q_{[\mu}Q_{\nu}Q_{\rho]}\Big)\;, 
 \eea
which transforms under Eq.~\eqref{Qgauge} as 
 \bea\label{CStransform}
  \delta_{\Lambda}{\rm CS}_{\mu\nu\rho}(Q)=\partial_{[\mu}{\rm Tr}\big(\partial_{\nu}\Lambda\, Q_{\rho]}\big)\;. 
 \eea
The barred formulas are analogous. 
Working out the Chern-Simons-form for Eq.~\eqref{explicitQ} one finds precisely Eq.~\eqref{Omegafirst}, up to a global factor 3.  
We thus define the 3-form curvature with Chern-Simons modification as
\begin{equation}
   \label{newHH}
  \widetilde{H}_{\mu\nu\rho}\equiv H_{\mu\nu\rho} -\frac{3}{2}\, \alpha'\left({\rm CS}_{\mu\nu\rho}(Q)-  {\rm CS}_{\mu\nu\rho}(\bar{Q})\right)\;. 
\end{equation}
In this formulation the \Odd invariance is manifestly realized without deformation. Rather, the $\GL(d)\times \GL(d)$ gauge symmetry is deformed, 
with the 2-form transforming according to the Green-Schwarz mechanism,  
 \begin{equation}
  \delta B_{\mu\nu} = \frac{1}{2}\,\alpha'\,{\rm Tr}\big(\partial_{[\mu}\Lambda\, Q_{\nu]}\big)-\frac{1}{2}\,\alpha'\,{\rm Tr}\big(\partial_{[\mu}\bar{\Lambda}\, \bar{Q}_{\nu]}\big) \,. 
\end{equation}
Upon partial gauge fixing to ${\rm SO}(d)\times {\rm SO}(d)$ and field redefinition this transformation relates to the reduction
of $\alpha'$-deformed double field theory \cite{Baron:2017dvb}.
If we fully gauge fix $\GL(d)\times \GL(d)$ (as done above in order 
to obtain explicit forms like Eq.~(\ref{explicitQ})) the \Odd transformations get deformed through compensating gauge transformations so that $B_{\mu\nu}$ 
 starts transforming under \Odd. 

We finally discuss the $\mathbb{Z}_{2}$ invariance (\ref{Z2metric}) of bosonic string theory in the frame formulation. It acts on the frame field as 
 \begin{equation}\label{ZonFrame}
  E\rightarrow Z^T E \tilde{Z}\;, \quad \tilde{Z}\equiv \begin{pmatrix}
                    0 & 1 \\
                    1 & 0
                  \end{pmatrix}. 
 \end{equation}
The matrix $\tilde{Z}$ has the effect of exchanging the two GL$(d)$ factors or, equivalently, to exchange the role of unbarred and barred indices. 
Indeed, a quick computation shows that under Eq.~(\ref{ZonFrame}) the Maurer-Cartan forms (\ref{MC}) transform as $P_{\mu}\leftrightarrow \bar{P}_{\mu}$ 
and $Q_{\mu}\leftrightarrow \bar{Q}_{\mu}$, in agreement with the explicit form (\ref{explicitQ}). 
Thus, thanks to the relative sign in Eq.~(\ref{newHH}) the total Chern-Simons form is $\mathbb{Z}_{2}$ odd, which together with $B_{\mu\nu}\rightarrow -B_{\mu\nu}$
implies $\mathbb{Z}_{2}$ invariance.

We have shown that Green-Schwarz type mechanisms may be necessary  even in classical string theory, which includes $\alpha'$ corrections, 
in order to  realize its global duality symmetries. 
While the original Green-Schwarz mechanism of Ref.~\cite{Green:1984sg} does modify the classical theory, 
in that case the need for an $\alpha'$-deformation of gauge transformations follows by requiring cancellation of quantum anomalies.
The observation that already invariance of the purely classical theory 
requires Green-Schwarz type mechanisms 
suggests that these 
may play a much more general role, perhaps  in order to revisit U-duality in the presence of higher order corrections. 

\noindent\textbf{Acknowledgements}:  
We thank J. Maharana and D. Marqu\'es for helpful discussions.
The work of O.H. is supported by the ERC Consolidator Grant ``Symmetries \& Cosmology''.


\begin{thebibliography}{10}

\bibitem{EloyHohmSamtleben}
C.~Eloy, O.~Hohm, and H.~Samtleben. in preparation.

\bibitem{Sen:1991zi}
A.~Sen, ``{$O(d) \otimes O(d)$} symmetry of the space of cosmological solutions
  in string theory, scale factor duality and two-dimensional black holes,''
\href{http://dx.doi.org/10.1016/0370-2693(91)90090-D}{{\em Phys. Lett.}
  {\bfseries B271} (1991) 295--300}.

\bibitem{Veneziano:1991ek}
G.~Veneziano, ``Scale factor duality for classical and quantum strings,''
\href{http://dx.doi.org/10.1016/0370-2693(91)90055-U}{{\em Phys. Lett.}
  {\bfseries B265} (1991) 287--294}.

\bibitem{Maharana:1992my}
J.~Maharana and J.~H. Schwarz, ``Noncompact symmetries in string theory,''
  \href{http://dx.doi.org/10.1016/0550-3213(93)90387-5}{{\em Nucl.Phys.}
  {\bfseries B390} (1993) 3--32},
\href{http://arxiv.org/abs/hep-th/9207016}{{\ttfamily arXiv:hep-th/9207016
  [hep-th]}}.

\bibitem{Meissner:1996sa}
K.~A. Meissner, ``Symmetries of higher order string gravity actions,''
  \href{http://dx.doi.org/10.1016/S0370-2693(96)01556-0}{{\em Phys. Lett.}
  {\bfseries B392} (1997) 298--304},
\href{http://arxiv.org/abs/hep-th/9610131}{{\ttfamily arXiv:hep-th/9610131
  [hep-th]}}.

\bibitem{Kaloper:1997ux}
N.~Kaloper and K.~A. Meissner, ``Duality beyond the first loop,''
  \href{http://dx.doi.org/10.1103/PhysRevD.56.7940}{{\em Phys. Rev.} {\bfseries
  D56} (1997) 7940--7953},
\href{http://arxiv.org/abs/hep-th/9705193}{{\ttfamily arXiv:hep-th/9705193
  [hep-th]}}.

\bibitem{Godazgar:2013bja}
H.~Godazgar and M.~Godazgar, ``{Duality completion of higher derivative
  corrections},'' \href{http://dx.doi.org/10.1007/JHEP09(2013)140}{{\em JHEP}
  {\bfseries 1309} (2013) 140},
\href{http://arxiv.org/abs/1306.4918}{{\ttfamily arXiv:1306.4918 [hep-th]}}.

\bibitem{Siegel:1993th}
W.~Siegel, ``{Superspace duality in low-energy superstrings},''
  \href{http://dx.doi.org/10.1103/PhysRevD.48.2826}{{\em Phys.Rev.} {\bfseries
  D48} (1993) 2826--2837},
\href{http://arxiv.org/abs/hep-th/9305073}{{\ttfamily arXiv:hep-th/9305073
  [hep-th]}}.

\bibitem{Hull:2009mi}
C.~Hull and B.~Zwiebach, ``Double field theory,''
  \href{http://dx.doi.org/10.1088/1126-6708/2009/09/099}{{\em JHEP} {\bfseries
  0909} (2009) 099},
\href{http://arxiv.org/abs/0904.4664}{{\ttfamily arXiv:0904.4664 [hep-th]}}.

\bibitem{Hohm:2010pp}
O.~Hohm, C.~Hull, and B.~Zwiebach, ``{Generalized metric formulation of double
  field theory},'' \href{http://dx.doi.org/10.1007/JHEP08(2010)008}{{\em JHEP}
  {\bfseries 1008} (2010) 008},
\href{http://arxiv.org/abs/1006.4823}{{\ttfamily arXiv:1006.4823 [hep-th]}}.

\bibitem{Hohm:2013jaa}
O.~Hohm, W.~Siegel, and B.~Zwiebach, ``Doubled $\alpha'$-geometry,''
  \href{http://dx.doi.org/10.1007/JHEP02(2014)065}{{\em JHEP} {\bfseries 1402}
  (2014) 065},
\href{http://arxiv.org/abs/1306.2970}{{\ttfamily arXiv:1306.2970 [hep-th]}}.

\bibitem{Hohm:2014eba}
O.~Hohm and B.~Zwiebach, ``{Green-Schwarz} mechanism and $\alpha'$-deformed
  {C}ourant brackets,'' \href{http://dx.doi.org/10.1007/JHEP01(2015)012}{{\em
  JHEP} {\bfseries 01} (2015) 012},
\href{http://arxiv.org/abs/1407.0708}{{\ttfamily arXiv:1407.0708 [hep-th]}}.

\bibitem{Hohm:2014xsa}
O.~Hohm and B.~Zwiebach, ``Double field theory at order $\alpha'$,''
  \href{http://dx.doi.org/10.1007/JHEP11(2014)075}{{\em JHEP} {\bfseries 11}
  (2014) 075},
\href{http://arxiv.org/abs/1407.3803}{{\ttfamily arXiv:1407.3803 [hep-th]}}.

\bibitem{Marques:2015vua}
D.~Marques and C.~A. Nunez, ``{T}-duality and $\alpha'$-corrections,''
  \href{http://dx.doi.org/10.1007/JHEP10(2015)084}{{\em JHEP} {\bfseries 10}
  (2015) 084},
\href{http://arxiv.org/abs/1507.00652}{{\ttfamily arXiv:1507.00652 [hep-th]}}.

\bibitem{Hohm:2016lge}
O.~Hohm, ``Background independence and duality invariance in string theory,''
  \href{http://dx.doi.org/10.1103/PhysRevLett.118.131601}{{\em Phys. Rev.
  Lett.} {\bfseries 118} no.~13, (2017) 131601},
\href{http://arxiv.org/abs/1612.03966}{{\ttfamily arXiv:1612.03966 [hep-th]}}.

\bibitem{Hohm:2016yvc}
O.~Hohm, ``Background independent double field theory at order $\alpha'$:
  {M}etric vs. frame-like geometry,''
  \href{http://dx.doi.org/10.1103/PhysRevD.95.066018}{{\em Phys. Rev.}
  {\bfseries D95} no.~6, (2017) 066018},
\href{http://arxiv.org/abs/1612.06453}{{\ttfamily arXiv:1612.06453 [hep-th]}}.

\bibitem{Baron:2017dvb}
W.~H. Baron, J.~J. Fernandez-Melgarejo, D.~Marques, and C.~Nunez, ``The {Odd}
  story of $\alpha'$-corrections,''
  \href{http://dx.doi.org/10.1007/JHEP04(2017)078}{{\em JHEP} {\bfseries 04}
  (2017) 078},
\href{http://arxiv.org/abs/1702.05489}{{\ttfamily arXiv:1702.05489 [hep-th]}}.

\bibitem{Green:1984sg}
M.~B. Green and J.~H. Schwarz, ``Anomaly cancellation in supersymmetric
  {$D=10$} gauge theory and superstring theory,''
\href{http://dx.doi.org/10.1016/0370-2693(84)91565-X}{{\em Phys. Lett.}
  {\bfseries 149B} (1984) 117--122}.

\bibitem{Metsaev:1987zx}
R.~R. Metsaev and A.~A. Tseytlin, ``Order $\alpha'$ (two loop) equivalence of
  the string equations of motion and the $\sigma$-model {W}eyl invariance
  conditions: {D}ependence on the dilaton and the antisymmetric tensor,''
\href{http://dx.doi.org/10.1016/0550-3213(87)90077-0}{{\em Nucl. Phys.}
  {\bfseries B293} (1987) 385--419}.

\bibitem{Hohm:2015doa}
O.~Hohm and B.~Zwiebach, ``{T}-duality constraints on higher derivatives
  revisited,'' \href{http://dx.doi.org/10.1007/JHEP04(2016)101}{{\em JHEP}
  {\bfseries 04} (2016) 101},
\href{http://arxiv.org/abs/1510.00005}{{\ttfamily arXiv:1510.00005 [hep-th]}}.

\bibitem{Hohm:2019jgu}
O.~Hohm and B.~Zwiebach, ``Duality invariant cosmology to all orders in
  $\alpha'$,''
\href{http://arxiv.org/abs/1905.06963}{{\ttfamily arXiv:1905.06963 [hep-th]}}.

\end{thebibliography}

\providecommand{\href}[2]{#2}\begingroup\raggedright\endgroup

\end{document}